\def\year{2020}\relax
\documentclass[letterpaper]{article} %
\usepackage{aaai20}  %
\usepackage{times}  %
\usepackage{helvet} %
\usepackage{courier}  %
\usepackage[hyphens]{url}  %
\usepackage{graphicx} %
\usepackage{graphicx}  %
\nocopyright %

\title{On Memory Mechanism in Multi-Agent Reinforcement Learning}
\author{Yilun Zhou,\textsuperscript{\rm 1} Derrik E. Asher,\textsuperscript{\rm 2} Nicholas R. Waytowich,\textsuperscript{\rm 2} Julie A. Shah\textsuperscript{\rm 1}\\ %
\textsuperscript{\rm 1}MIT CSAIL, \textsuperscript{\rm 2}US Army Research Lab\\ 
yilun@csail.mit.edu
}

\usepackage{amsmath}
\usepackage{amssymb}
\usepackage{bbm}
\usepackage{booktabs}
\usepackage{natbib}

\DeclareMathOperator*{\argmax}{arg\,max}

\let\vec\mathbf

\begin{document}

\maketitle

\begin{abstract}
Multi-agent reinforcement learning (MARL) extends (single-agent) reinforcement learning (RL) by introducing additional agents and (potentially) partial observability of the environment. Consequently, algorithms for solving MARL problems incorporate various extensions beyond traditional RL methods, such as a learned communication protocol between cooperative agents that enables exchange of private information or adaptive modeling of opponents in competitive settings. One popular algorithmic construct is a memory mechanism such that an agent's decisions can depend not only upon the current state but also upon the history of observed states and actions. In this paper, we study how a memory mechanism can be useful in environments with different properties, such as observability, internality and presence of a communication channel. Using both prior work and new experiments, we show that a memory mechanism is helpful when learning agents need to model other agents and/or when communication is constrained in some way; however we must to be cautious of agents achieving effective memoryfulness through other means. 
\end{abstract}

\section{Introduction}
Compared to (single-agent) reinforcement learning (RL), multi-agent reinforcement learning (MARL) includes other agents and potentially partial observability into the environment. Due to the partial observability of an environment from each agent, emergent communication \citep{grounded-comm, commnet, rial, lang-learning-paradigm}, in which agents must come up with a communication protocol to efficiently exchange information, has been studied. Another learning objective is for a few agents to learn the policies and then get deployed in an environment with other agents. We call these learning agents as \em internal \rm agents, while the rest are called \em external \rm agents. In this setting, internal agents should be adaptive to the different behaviors exhibited by external agents, regardless of whether the agents are in cooperative, competitive, or mixed relations.

A recurring theme within MARL algorithms is the use of a memory mechanism, which is most commonly implemented by a recurrent neural network for deep learning methods, or sometimes by state augmentation if the system designer determines that a bounded memory is sufficient. However, sometimes the use of such a construct is not well-motivated, with both memoryful and memoryless policy representations having been tested in prior work \citep{commnet}. Furthermore, there has not been a systematic study of whether or how such a memory mechanism can be helpful. In this paper, we group multi-agent environments according to three properties, and study the benefit that a memory mechanism may provide. Throughout the study, we ground our propositions with conclusions derived from existing literature and new experiments, and consider both theoretical and practical feasibility. 

At a high level, we classify MARL environments along three axes (detailed in Section \ref{sec-taxonomy}). First, we consider the observability axis, or whether each agent is able to receive the full state information at each time. Several simple benchmarks, such as the predator-prey task and the cooperative navigation task introduced by \citet{maddpg}, are fully observable. As special cases, classical game theory examples (such as rock-paper-scissors and prisoners' dilemma) are also fully observable since they are stateless. In stark contrast, a common source of non-observability comes from the limited sensing range of agents, as in the traffic junction domain used by \citet{commnet}. 

Second, we consider the internality of the environment, or whether the environment contains only internal agents (over which we have control \em during test time\rm), or also includes external agents (which are provided by other parties, \em during test time\rm). If all agents are internal (referred to as \em fully internal \rm environments), all agents are essentially optimizing for a shared reward within a fully cooperative setting. In \em partially internal \rm environments, agents may be engaged in fully cooperative tasks (e.g., human-robot teaming for rescue), fully competitive tasks (e.g., a predator-prey scenario), or mixed tasks (e.g., traffic, in which vehicles compete for road resources while also cooperating to avoid collision). 

Third, we consider an axis for communication in fully internal environments: specifically, whether an explicit communication channel is present. If a channel is present, agents can learn to exchange information effectively, which may lead to improved performance. 

In this work, we study the utility of a memory mechanism for environments according to the above three axes, and report the following: 
\begin{enumerate}
    \item Within partially internal environments, regardless of reward structure, memoryful policies are helpful because they can better model opponents and inherently adapt to potentially unseen opponent behaviors. 
    \item Although the presence of a communication channel without restrictions does not necessitate a memory mechanism, it can be helpful for agents to evolve compositional and/or sparse communications. 
    \item In fully internal environments without explicit communication, a memory mechanism can typically help for agents to engage in behavior-based communication. 
    \item An emergent phenomenon, which we refer to as ``state-based bookkeeping'', can sometimes mitigate or remove the duty of an explicit memory mechanism, which may or may not be desirable. 
\end{enumerate}

\section{Background}

Similar to MDP representations in RL, a MARL environment with $n$ agents can be specified by a tuple $\langle S, A_1, ..., A_n, T, R_1, ..., R_n, \gamma\rangle$. $S$ represents the state space (which includes the state of all agents and additional environmental information). $A_i$ denotes the set of actions for agent $i$.  $T: S\times A_1 \times ... \times A_n\rightarrow \mathbb P_S$ represents the transition function, assuming that agents actions are simultaneous rather than occurring in turns. $R_i: S\times A_1 \times ... \times A_n \times S \rightarrow \mathbb R$ represents the reward of agent $i$, which is a function of the current state, all agents' actions, and the next state. $\gamma \in [0, 1]$ denotes the discount factor. 

The optimization objective is dependent upon environment settings. For example, if agents are fully cooperative, the goal is to identify policies that maximize the total expected discounted reward, $\mathbb E[\sum_{t=1}^\infty \gamma^t \sum_{i=1}^n R_i(s_t, \pi_1(s_t), ..., \pi_n(s_t), T(s_t, \pi_1(s_t), ..., 
\allowbreak \pi_n(s_t))]$. In this case, we can also formulate the reward functions to be identical and equal to the average of all individual rewards (i.e., $\bar R=\sum_{i=1}^n R_i / n$ represents the new reward that each agent is maximizing individually). This is the formulation used in most literature on cooperative agents \citep{commnet, rial}. 

A selfish agent trying to maximize its own reward may want to guarantee a worst-case optimal reward, resulting in a policy that receives the highest reward regardless of other agents' policies. Alternatively, the selfish agent may want to achieve an average-case optimal policy, which yields the highest expected reward assuming some distribution of opponent policy. 

Notably, in contrast to single-agent RL, a deterministic policy is often not the best policy (and can even be the worst). For example, in an iterative version of the rock-paper-scissors game, any deterministic policy is exploitable and should result in repetitive losses. The worst-case optimal policy is a perfectly randomized policy that achieves equal proportion of win, loss, and draw in expectation. 

\section{Memoryful Policies}
It is known \citep{suttonbarto} that the optimal policy in an MDP is Markov. In other words, choosing (distribution of) actions based only on the current state can lead to the optimal policy, and maintaining state and/or action history is not further advantageous. However, due to the presence of other agents, it may be helpful for the policy to be memoryful. As a simple example, in a repeated rock-paper-scissors game, against an imperfectly playing agent with a tendency toward a specific action, a memoryful agent can exploit opponent imperfection by calculating the probability of each action and taking corresponding counter-actions. Note that this kind of ``learning'' is performed on the fly, within each episode, and thus will adapt to a distinctly different opponent next time without issue. 

We define a memoryful policy for the $i$-th agent in the general form as $\pi_i: \prod_{t=1}^\tau S\times\prod_{j=1}^n\prod_{t=1}^{\tau-1} A_j\rightarrow A_i$. In terms of implementation, a general memoryful policy can be represented by a recurrent neural network. However, if only a certain length of past history is necessary, state augmentation technique can also be used, as in the state representation used by \citet{lola} in the iterative prisoners' dilemma. 

\section{A Taxonomy of MARL Problems}
\label{sec-taxonomy}
For the purpose of this work, we classify MARL problems in terms of three attributes, each of which has different implications on the requirement for memory mechanisms. 

\begin{enumerate}
    \item Observability: whether each agent \em individually \rm receives the full state information of the environment, or only a correlated observation. Partial observability is more common in multi-agent systems, but full observability is typical in many game theory domains (such as rock-paper-scissors or the prisoners' dilemma). 
    
    \item Internality: whether all agents are internal, or some external agents exist. Specifically, as discussed in the introduction, we define an internal agent to be one that we can control at \em test \rm time, while an external agent is supplied by another party during \em test \rm time. If all agents are internal, we refer to that environment as \em fully internal\rm; otherwise the environment is \em partially internal\rm. We assume a perfectly shared reward structure for fully internal environments (i.e. fully cooperative), because any source of competition would come from external agents in realistic situations\footnote{Note that despite rewards being shared, the training algorithm is not restricted to using the same reward for all agents: it can reward agents that perform the bulk of the work while penalizing non-participating agents. The method credit assignment has received much attention \citep{counterfactual-grad, credit-assignment}; however, for the sake of simplicity, we do not consider it in this paper. }. 
    
    Note that partial internality does not necessarily imply a competitive setting. For example, in human-robot teaming for search and rescue, the reward is still shared. Rather, it is simply that the robot must view humans as external. A traffic junction represents a mixed cooperative/competitive environment, since vehicles must compete for road usage while cooperating to avoid collisions. 
    
    \item Communication: whether explicit message-based communication is allowed \em among internal agents\rm. Specifically, the directed communication channel from agent $i$ to $j$ is represented by a function $c_{ij}: \mathbb R^{d_i}\rightarrow \mathbb R^{d_j}$, where agent $i$ can send a $d_i$-dimensional message $m_{ij}$, and agent $j$ will receive an encoded version $m_{ij}' = c_{ij}(m_{ij})\in \mathbb R^{d_j}$. Typically, the channel encoder is the identity function $c_{ij}(m)=m$, but the encoder can also be a discretization function to enforce discrete communication signals \citep{grounded-comm, rial} or a function that suppresses the signal after a certain number of times to implement budgeted communication \citep{r-maddpg}. We assume that for agent $i$ the outbound communication message $m_{ij}$ is computed from individual state observation, and the action is computed from both the state observation and inbound communication messages $m_{ki}'$. We do not consider a communication channel between internal and external agents in this work. 
\end{enumerate}

\section{Fully Observable/Fully Internal}
For fully observable environments with fully internal agents, the optimal policy is still memoryless (i.e. Markovian), at least in theory. Specifically, we can treat the problem as single-agent RL in which an agent learns a policy that maps from the state space to the joint action space. Since the environment is represented by a MDP, the optimal policy for this agent is also Markovian (and deterministic). At test time in the MARL problem, then, each individual agent calculates the joint action using the original policy, and takes its individual action correspondingly. Communication is also not required in this case, because the learned joint policy contains any necessary coordination among agents. 

In practice, however, scalability can be an issue for environments containing too many agents, since the state space and joint action spaces will grow exponentially with an increasing number of agents. Therefore, intentional decentralization can be helpful: with such a formulation, each agent only observes local state information. Moreover, given homogeneous agents and parameter sharing, learned policies also have the potential to adapt to an indefinite number of agents \citep{commnet}. In this case, depending on requirements for communication, a memory mechanism can be beneficial. We discuss this in greater detail in Section \ref{sec-partobs-fullint}. 

\section{Fully Observable/Partially Internal}
As noted in the prior section, a collection of internal agents can function effectively as one agent at both training and test time assuming full observability: thus, we only consider the case of a single internal agent deployed into an environment with other external agents. In addition, we use the word ``opponent'' to refer to the other agent, even in partially or fully cooperative situations. 

There has been a significant amount of research on learning competitive behaviors in game-theoretic settings. The methods for doing so can be classified according to the ways they model opponents. 

The first way to address opponent modeling is to simply not model them. Obviously this is potentially problematic and vulnerable. However, \citet{minimax-q} proposed a method to learn the worst-case optimal policy with a guaranteed reward against all possible type of opponents. That paper further indicates that such a policy is memoryless, but may be stochastic. In addition, \citet{m3ddpg} incorporated such worst-case consideration into a deep MARL algorithm proposed by \citet{maddpg}. 

The second approach is to model opponents as static or slowly changing, so that agents can learn to track opponent policies during test time. Most recent deep MARL approaches have adopted this assumption and use co-learning at training time, which simulates an opponent policy and learns to better deal with that policy, while simultaneously updating the opponent policy in order to be stronger in response. For example, \citet{maddpg} proposed a general-purpose MARL algorithm that can learn from arbitrary reward structures. \citet{lola} added a lookahead step to affect opponent gradient update with opponent modeling. \citet{meta-marl} used meta-learning for fast adaptation with respect to opponent evolution. Although these methods do not explicitly require a memory mechanism, more interesting behaviors can emerge with some form of memory, as demonstrated by \citet{lola}, who used state augmentation in their iterated prisoners' dilemma setting. 

\begin{figure}[!htb]
    \centering
    \includegraphics[width=0.8\columnwidth]{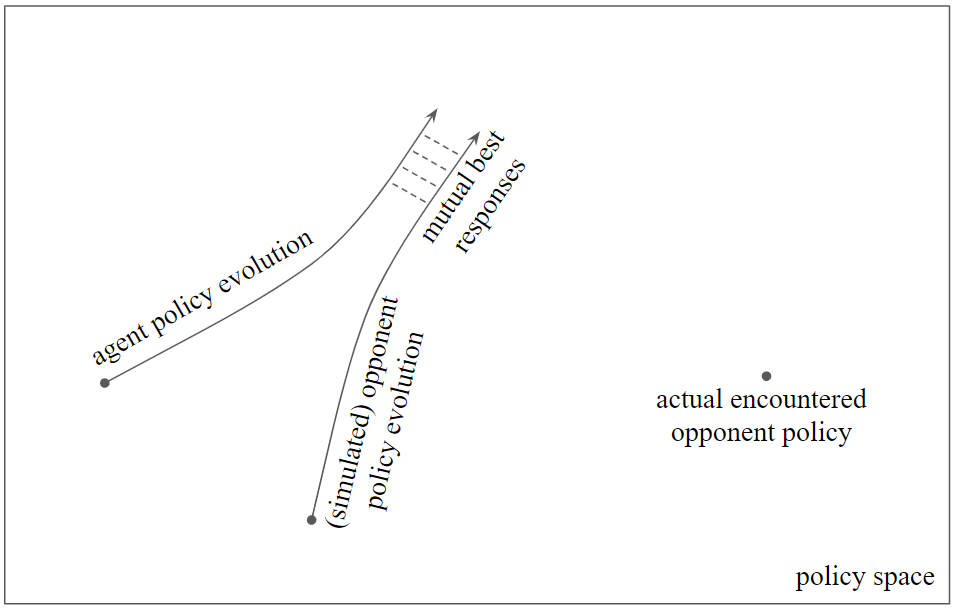}
    \caption{Training dynamics of co-learning algorithms. }
    \label{convergence}
\end{figure}

However, when the opponent is using a very different policy from the one modeled during training, or is changing policy abruptly across episodes, the agent may receive very bad rewards. Figure \ref{convergence} sketches a possible failure case for this kind of co-learning based method. In fact, many co-learning approaches lack a guarantee that the current agent policy remains effective against a previous opponent policy, which is analogous to the problem of forgetting in both discriminative \citep{nn-forgetting} and generative \citep{gan-forgetting} models. 

In addition, \citet{measurement-pitfalls} found that in policies trained using such co-learning methods, while positive signaling is present, positive listening is not. In other words, the receiving agent does not seem to base its actions upon the sending agent's communication. This suggests that, in a stateless environment such as a matrix game with communication, the receiving agent's policy implicitly learns the best response to the sending agent's policy (with which the message is correlated, as observed with positive signaling), but would poorly adapt to any policy change by the sending agent, even with intention signaling. 

The last final approach, which has been relatively less studied, is to model opponents explicitly at test time. Specifically, \em within an episode\rm, an explicit memory mechanism is used to simultaneously infer and adapt to the current opponent policy. This method places relatively few assumptions on the opponent and can adapt to drastic changes to opponent policies across episodes. 

Theory of mind \citep{tom} is a popular framework for such reasoning. For example, \citet{tomnet} used a mental state network to model opponent policy from past behavior. \citet{som} learned to infer the goals of opponent in an online manner and correspondingly adapt the agent's own behavior. \citet{bayes-tomop} assumed switching behaviors among several stationary policies for the opponent and inferred the acting policy. 

\begin{figure}[!htb]
    \centering
    \includegraphics[width=0.45\columnwidth]{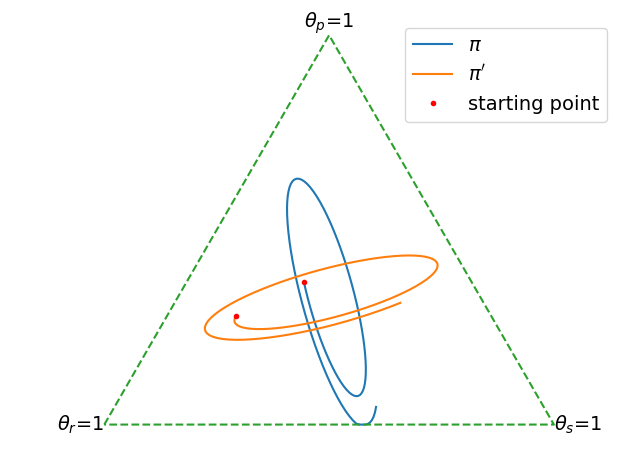}
    \includegraphics[width=0.45\columnwidth]{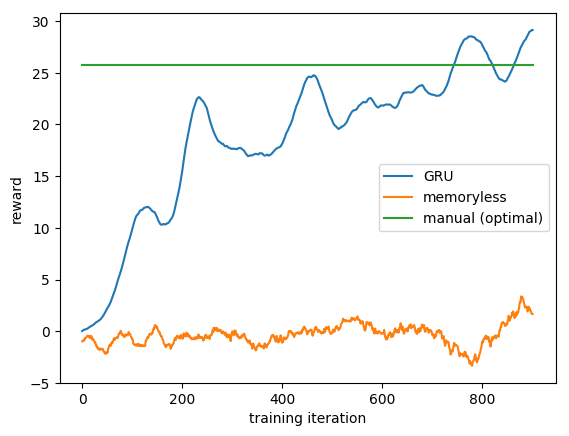}
    \caption{Left: trajectories of agent and opponent policies with a co-learning approach exhibit oscillation and non-convergence. Right: agent reward for the repeated RPS game, with uniformly distributed opponents. }
    \label{rps-curve}
\end{figure}

Another advantage is that this method can exploit any imperfection or weakness of a given opponent. To see this, consider again the (stateless) game of iterated rock-paper-scissors, wherein the policy is simply parametrized by the probabilities of outputting rock, paper, or scissors ($\pi=[\theta_r, \theta_p, \theta_s]$). A worst-case optimal method \citep{minimax-q} ignoring the opponent's policy would compute the policy to be $\pi=[1/3, 1/3, 1/3]$, with an expected reward of 0. Co-learning algorithms would produce a pair of policies that oscillate on this probability simplex. Specifically, for two agent policies parametrized by $\pi$ and $\pi'$, the expected average reward for the $\pi$ agent is as follows:
\begin{align*}
    \mathbb E[R_{\pi, \pi'}]=\theta_r\theta'_s+\theta_p\theta'_r+\theta_s\theta'_p-\theta_r\theta'_p-\theta_p\theta'_s-\theta_s\theta'_r. 
\end{align*}
For the $\pi'$ agent in this case, the expected average reward is $-\mathbb E[R_{\pi, \pi'}]$. 
Thus, the policy gradient is $(\theta'_s-\theta'_p, \theta'_r-\theta'_s, \theta'_p-\theta'_r)$ for $\pi$, and $(\theta_s-\theta_p, \theta_r-\theta_s, \theta_p-\theta_r)$ for $\pi'$. The left diagram of Figure \ref{rps-curve} dipicts the oscillation of the policy update.

Nevertheless, at any given time, if the learned agent is deployed in an environment, even with an opponent whose policy is uniformly sampled from this simplex, the expected reward is 0. Use $\pi$ to denote the agent policy, and $\pi'$ to denote the opponent policy. The probability density function for the uniform opponent policy is $p(\pi')=\sqrt 3/2\cdot \mathbbm 1_{\theta'_r+\theta'_p+\theta'_s=1}$. Thus $\mathbb E_{\pi'} [\mathbb E[R_{\pi, \pi'}]]=0$ by symmetry. 

By comparison, a simple memoryful policy would estimate the opponent policy parameters by action history, and take the corresponding counter-action, as implemented by the following hand-crafted recurrent neural network (rock, paper, and scissors have the index of 0, 1, and 2 respectively). 
\begin{align*}
    \vec h_0 &= [0, 0, 0],\\
    \vec h_t &= \gamma \vec h_{t-1} + \vec x_t,\\
    i_t &= \argmax \vec h_t,\\
    y_t &= [1, 2, 0][i_t]. 
\end{align*}
With $\gamma=1$, $\vec h_t$ represents the count for opponent actions, $\vec x_t$ the current opponent action, and $y_t$ the current agent action. If we expect the opponent policy to be not stationary, a discount factor $\gamma\leq 1$ can be added in order to favor more recent opponent actions. 

The simple procedure exemplifies the idea that using a memory mechanism to explicitly model opponents and derive belief about the opponent policy ($h_t$), upon which the corresponding decision is made ($y_t$). In some sense, this idea is similar to the RL$^2$ approach proposed by \citet{rl2} in which variations of a single agent RL environment are learned. 

Ideally, the recurrent calculation is learned from data with general-purpose function approximators such as LSTM \citep{lstm} or GRU \citep{gru} cells. For example, Figure \ref{rps-curve} plots the reward for training a GRU network with a policy gradient, along with the average performance for the above manually specified policy (which is statistically optimal). Our learned policy is able to reach the optimal policy within hundreds of training episodes, while a memoryless policy based on the co-learning approach is unable to learn against drastically changing opponents across episodes. 

A more real-life domain is autonomous driving. Human driver modeling is an important aspect of it, which is why pure reinforcement learning approaches are not very common, but often combined with supervised or imitation learning from human driving data \citep{supervised-driving, imitation-driving}. In addition, co-learning driving policies are not likely to be successful during deployment because the learned opponent behavior may not be similar to actual human behaviors\footnote{Such observation is captured by the argument that autonomous vehicles (AVs) driving alongside other AVs is actually easier than driving alongside human drivers.}. 

\begin{figure}[!htb]
    \centering
    \includegraphics[width=\columnwidth]{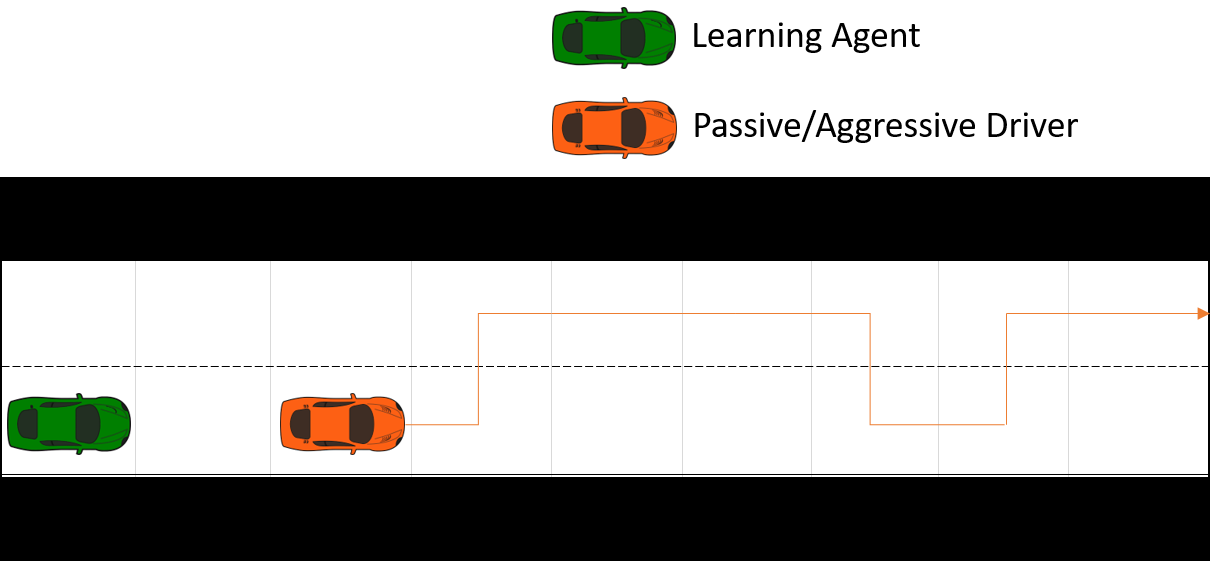}
    \caption{Traffic domain. }
    \label{traffic-gridworld}
\end{figure}

As a concrete example, consider driving on a two-lane road with a leading vehicle in front, represented as a grid world as shown in Figure \ref{traffic-gridworld}. At each time step, each vehicle decides to either move forward (right) one position, change lane, or stay still. Our agent is rewarded for fast traversal through the road while avoiding collision. Therefore, its optimal policy for our agent depends on the other agent: if the other agent is driving slowly, our agent may wish to pass it on another lane, but not if the other vehicle is changing lanes erratically. Specifically, we model four types of behavior for the other agent, whose action probabilities are summarized in Table \ref{behavior-prob}. 
\begin{table}[!htb]
    \centering
    \begin{tabular}{c|c|c|c}\toprule
        type & forward & lane-change & stay \\\midrule
        passive-fast (PF) & 90\% & 10\% & 0\% \\
        passive-slow (PS) & 20\% & 10\% & 70\%\\
        aggressive-fast (AF) & 30\% & 70\% & 0\%\\
        aggressive-slow (AS) & 20\% & 70\% & 10\%\\\bottomrule
    \end{tabular}
    \caption{Proportion of each action for different behaviors.}
    \label{behavior-prob}
\end{table}

Additionally, we consider three policies for our agent. The \em greedy \rm policy will always try to pass the other agent. The \em conservative \rm will always try to stay behind the other agent and avoid collision. Finally, the \em adaptive \rm policy will infer the other agent's policy using maximum-likelihood estimation and choose the greedy strategy for passive opponent and conservative strategy for aggressive opponent. 

With total road length of 30, time penalty of 1, crash penalty of 30, and initial separation of the two vehicles of 10 (opponent in front), Table \ref{traffic-reward} shows the reward of the three policies against the four type of behaviors, as well as a uniform mixture of behavioral types. 
\begin{table}[!htb]
    \centering
    \begin{tabular}{c|c|c|c|c|c}\toprule
        policy & PF & PS & AF & AS & Mix \\\midrule
        greedy & \textbf{-29.0} & -33.4 & -660.3 & -237.3 & -227.5\\
        conservative & \textbf{-29.0} & -94.3 & \textbf{-70.1} & \textbf{-94.5} & -71.1\\
        adaptive & \textbf{-29.0} & \textbf{-33.3} & -70.9 & -95.3 & \textbf{-57.3}\\\bottomrule
    \end{tabular}
    \caption{Reward for each policy against different behaviors. }
    \label{traffic-reward}
\end{table}

We can see that while memoryless greedy and conservative policies can each achieve high reward on different static opponent behaviors, both perform poorly if the opponent is randomizing its behavior per episode. On the other hand, the adaptive policy performs opponent modeling within each episode and can therefore achieve high reward regardless of opponent behaviors.

\section{Partially Observable/Fully Internal}
\label{sec-partobs-fullint}

In this section we assume that all agents have joint access to the full state information for a given environment. If all agents (combined) do not receive the full state information, this problem is at least as hard as a partially observable Markov decision process (POMDP) \citep{pomdp}, and a memory mechanism is necessary (e.g. as done recently with deep recurrent Q-learning \citep{drqn}). 

\subsection{With (explicit) communication}
Much recent research has been focused on the use of communication to complete cooperative tasks (mostly through reinforcement learning \citep{lang-learning-paradigm}). For example, \citet{commnet} provided an end-to-end model for learning agent communication between an indefinite amount of agents. \citet{rial} proposed a differentiable communication channel so that back-propagation could be performed across time steps. \citet{maddpg} demonstrated  that arbitrary communication channels can be incorporated. None of the above works made use of a memory mechanism, since the channel capacity is greater than that required to encode the necessary information exchange. %

In general, if the communication channel capacity is sufficient and agents can jointly observe the full state information, they can exchange such information through communication. This is the setting studied in most literature related to learning communication strategies \citep{maddpg, commnet}. Since actions will be taken upon receiving the messages, a memory mechanism would not be helpful in such scenario. 

However, if the communication channel can only output a finite and discrete set of tokens, a memory mechanism is necessary in order to learn a compositional semantic of messages. For example, \citet{grounded-comm} found that interpretable compositional language can emerge from cooperative memoryful agents. \citet{symbol-referral} conducted a similar study, but with the target of learning to refer to pictures. \citet{visual-dialog} showed that dialog communication can also emerge in a similar image guessing game. \citet{multi-step-game} studies a similar variant of image referral game by incorporating multi-modality. \citet{obverter-referral} and \citet{obverter-gridworld} both used the obverter technique \citep{obverter-original} to learn language, based on the assumption that the model for the receiving agent is the same as that for the sending agent, but for different environments: a grid world environment and a referral game. Without exception, all these works implemented recurrent policies. 

Another constraint is if the channel only allows a certain amount of messages to pass before being non-transmitting. This circumstance requires a memory mechanism on both the sender and receiver ends to memorize the last message being exchanged, so that limited amount of communication attempts can be put into best possible use. \citet{r-maddpg} confirm that agents are unlikely to learn good policies in the absence of such mechanism. 

\subsection{Without (explicit) communication}
In the absence of explicit communication, agents can still find ways to communicate their specific behaviors. This is perhaps unsurprising, because such behavior-based communication is also common among human interactions, with body language serving as a prominent example. However, a memory mechanism is typically required to both generate and parse communicative behaviors, which are temporally extended. 

\begin{figure}[!htb]
    \centering
    \includegraphics[width=0.95\columnwidth]{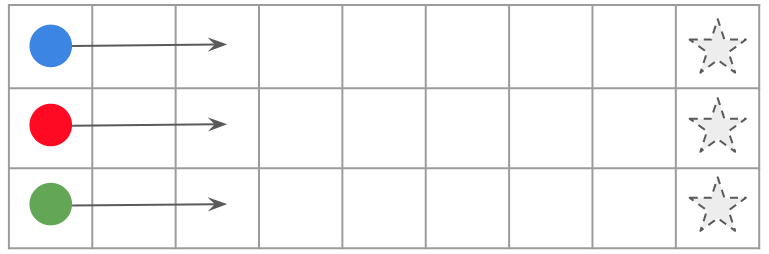}
    \caption{The environment for the three-lane target pursuit task, without explicit communication.}
    \label{3-lane}
\end{figure}

As an illustrative example, consider the environment depicted in Figure \ref{3-lane}. Three agents start at the left of three lanes, while a target will appear randomly on the right of one lane. The agents are not allowed to cross lanes and reward is received when the target is reached. At each time step, each agent can choose to move to the left or right for one position, or stay still, and will incur a cost for movement (e.g. fuel cost). Only the center (red) agent can observe the target location, but all agents can observe other agents. 

Without cost for agent movement, the optimal policy for this problem is for all three agents to independently head off to pursue the potential target (with only one agent ultimately arriving at that target). However, in the presence of such a cost, a better policy would be for the center agent to communicate target's location \em with its behavior\rm, which is in turn parsed by the two other agents to receive the corresponding message. 

\begin{table}[!htb]
    \centering
    \begin{tabular}{c|c}\toprule
        target location & first two steps \\\midrule
        top & $\rightarrow, \leftarrow$\\
        center & $\rightarrow, \rightarrow$\\
        bottom & $\rightarrow, \cdot$\\
    \end{tabular}
    \caption{Example of behavior-based communication to solve the domain in Figure \ref{3-lane}.}
    \label{behavior-comm}
\end{table}

More concretely, consider the behavior-based communication of the red agent during the first two time steps, as shown in Table \ref{behavior-comm}. The red agent uses a RIGHT-LEFT behavior sequence to convey that the target is in the top lane, a RIGHT-NOP sequence to signal that the the target is in the bottom lane, and a RIGHT-RIGHT sequence to indicate that the target is in the center lane. Given a memory mechanism, the blue and green agents are able to parse such temporally extended behavioral communication and take corresponding actions. If fuel cost is greater than time cost, it is easy to see that such a policy achieves better reward than the naive policy of every agent moving to the right at the same time. 

Since the signaling agent must trade off using behavior for either task completion or communication, any communication must be sparse \citep{r-maddpg}, and often temporally extended since primitive behaviors (i.e. actions) are not necessarily expressive enough \citep{grounded-comm}. Therefore, both the sending and receiving agents could benefit from an explicit memory mechanism.

\section{Partially Observable/Partially Internal}

In this case, we not only have external agents, but also have several internal agents, each of which cannot fully observe the environmental state. From the previous two sections, it was shown that a memory mechanism is needed to model opponents and to communicate with teammates under constraints. Therefore, a memory mechanism is also expected to help in this case, although it is studied by relatively few amount of work. For example, \citet{maddpg} proposed an algorithm that can work with this type of environment but is based on co-learning, which is prone to drastically changing external agents. %

\section{State-Based Bookkeeping}
Throughout this work, we assume that the agent achieves the goal of remembering past history only through the provided memory mechanism. However, this is not necessarily the case, as the agent can encode memory in unexpected places: specifically, the agent can store past history (i.e. opponent actions or teammate communication messages) within its own state. 

As an example in a discrete domain, consider a target-reaching task with cooperative communication in one-dimensional discrete space that includes four targets located at locations from 11 to 14. The mover agent starts at location 0, and must reach the correct pre-specified target (which is not observed), with a maximum movement of five grids each time. Its speaker teammate can observe the target location and send a discrete communication token from a set of two tokens $\{A, B\}$ each time, but cannot observe the mover's location. The episode is terminated when either of the four targets is reached, and reward of 10 is given for reaching the correct target. A cost of 1 is applied at each time step to encourage faster completion. 

Apparently, compositional language (i.e. phrases) is necessary to communicate the target location using the limited set of available tokens. In addition, when observing only the target as the state, the speaker requires a memory mechanism to generate a sequence of tokens. However, we show that the mover does not require an explicit memory mechanism in order to parse the communication. 

Table \ref{translation} depicts the speaker's emitted phrases corresponding to each target location, which we assume to be fixed. At each time step, the mover's policy is to move right a certain number of steps depending on its location and the received token at that time step: $a=\pi(s, c)$, where $c\in\{A, B\}$. 

\begin{table}[!htb]
    \centering
    \begin{tabular}{c|c}\toprule
        target location & message \\\midrule
        11 & $A, A$\\
        12 & $A, B$\\
        13 & $B, A$\\
        14 & $B, B$\\\bottomrule
    \end{tabular}
    \caption{Emitted message for each target location. Any token emitted beyond the 2nd time step is null. }
    \label{translation}
\end{table}

\begin{figure}[!htb]
    \centering
    \includegraphics[width=\columnwidth]{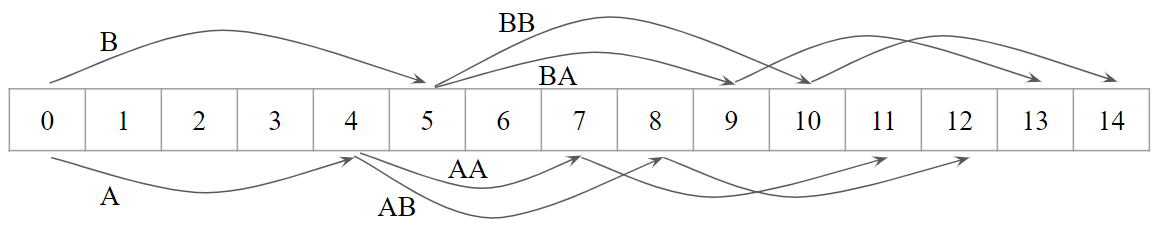}
    \caption{A memoryless policy incorporating state-based bookkeeping}
    \label{1d-bookkeeping-policy}
\end{figure}

Policy search finds the policy illustrated in Figure \ref{1d-bookkeeping-policy}. Upon receiving message token $A$, the agent moves to position 4, and moves to position 5 when receiving $B$. Therefore, its current position serves as a way to memorize the previous token. Then from position 4, it moves to 7 when receiving another $A$, and to 8 when receiving another $B$. Similarly, when from position 5, it moves to 9 when receiving $A$ and 10 when receiving $B$. Again, it uses its state to memorize the communication history (i.e. positions 7, 8, 9, and 10 clearly disambiguate messages $AA$, $AB$, $BA$, and $BB$). Then it will move a constant four steps to the desired target locations. Overall three time steps will be taken, which is also optimal even with a memory mechanism. Moreover, an optimal policy can be derived for any communication protocol as long as different targets are disambiguated by messages, facilitating the discovery of such solutions. 

Such a technique is also possible in continuous domains. In particular, \citet{r-maddpg} found that when using an actor-critic algorithm with a memoryful critic, memoryless actors are sufficient to achieve limited-budget communication, which requires agents to remember past messages: their finding may be explained by such phenomenon. Notably, similar phenomena have been discovered in other applications as well, such as neural network capable of hiding high-frequency information inside images \citep{gan-cheating}. Nevertheless, one way of mitigating such phenomenon is to make the transition stochastic, so that any high-frequency information would be destroyed by the transition function. 

In a similar vein, agents in communication-free environments can also employ this kind of state-based communication, as discussed above. However, unlike state-based memory mechanisms, such communications are to a large extent robust to environment noise, especially if such behaviors are temporally extended, with generation and parsing performed by a dedicated memory mechanism. However, when evaluating emergent communication protocols, it is critical to confirm whether agents are indeed using only the communicated messages or other agents' behaviors as well. 

\section{Conclusion and Future Work}
In this paper, we identified situations in multi-agent reinforcement learning can benefit from an explicit memory mechanism. The two major uses of such a mechanism is to model opponents and to carry out communicative behaviors. 

First, we argue that opponent modeling is an indispensable component of learning with external agents, regardless of the reward structure being shared or not. We found that the more popular co-learning approach employed by most algorithms will only keep an implicit model of the opponent and will poorly adapt to changing opponents, which is shown both experiments and confirmed by findings of \citet{measurement-pitfalls}. 

Second, we argue that a memory mechanism allows for more flexible communication protocols, both implicit and explicit. Most literature on communication focus on explicit communication, in which a dedicated communication channel is set up between the sending and the receiving agents. In this scheme, a memory mechanism has been shown to enable both temporally extended and sparse communications. In addition, we also show a memory mechanism can help emerge implicit, behavior-based communication, which is often temporally extended. 

Last, we identify a potential way for agents to bypass such a memory mechanism through what we call a state-based bookkeeping mechanism. Specifically, since state is persistently accessible by the agent during the Markovian transition, the agent may find a policy that uses state to implicitly encode memory content. 

One direction for future work is to design algorithms that can model more sophisticated opponents, including those that also adapt to our agent's behavior in an online manner (similar to \citep{lola} for co-learning algorithms). In addition, most studies of emergent communication is on explicit communication, but a memory mechanism also enables us to measure implicit communication. Moreover, the most general form of communication combines implicit and explicit ones, in that with a memory mechanism, the ``meaning'' as understood by the receiving agent is dependent not only on the explicit message, but also on the sending agent's movement, and potentially as well as the receiving agent's own state history. This may hinder interpretability, and it may be desirable to disentangle the contribution of each component to the perceived meaning of the message.

\bibliographystyle{aaai}
\bibliography{references}

\end{document}